\documentclass[preprint,showpacs,preprintnumbers,amsmath,amssymb]{revtex4}

\usepackage{graphicx}
\usepackage{dcolumn}
\usepackage{bm}

\begin{document}

\preprint{APS/123-QED}

\title{Experimental measurements of multiple stable trapped domain wall states induced in nanofabricated elements}

\author{D. Lacour}
\author{J. A. Katine}
\email{jordan.katine@hgst.com}
\author{L. Folks}
\author{T. Block}
\author{J. R. Childress}
\author{M. J. Carey}
\author{B. A. Gurney}

\affiliation{Hitachi San Jose Research Center, 650 Harry Rd, San
Jose CA 95120, USA.
}%

\date{\today}

\begin{abstract}
The presence of a domain wall trapped by a sub-micron notch is
probed in two ways: through electronic transport measurements and
by Magnetic Force Microscopy (MFM). We observe complex magnetic
features which are consistent with numerical simulations
predicting the existence of multiple magnetic configurations
stabilized by the notch structure.

\end{abstract}



\maketitle Domain walls (DW) play a critical role in much of the
current research involving spin-dependent electronic
devices\cite{Tatara99,Chopra02,Allwood02,Vernier03,Grollier02,Grollier03}.
In the case of a highly constrained DW, extremely large
magnetoresistance ratios have been
reported\cite{Tatara99,Chopra02}. Logic gates based on domain wall
motion are being studied\cite{Allwood02}.  Recently, the movement
of trapped DWs via the spin-transfer torque has been
experimentally demonstrated in thin
films\cite{Vernier03,Grollier02,Grollier03}. Although the DW
characteristics are key parameters in the aforementioned
experiments, only a few
theoretical\cite{Bruno99,McMichael00,Molyneux02} and experimental
studies\cite{Kaui03,Ono98} have been devoted to trapped DWs.

In this letter we present an experimental study of trapped DWs.
Using a device geometry previously proposed by Ono, et
al.\cite{Ono98} and subsequently employed by Grollier, et
al.\cite{Grollier02}, we demonstrate that multiple stable
locations exist for the trapped DW in such devices.

\section{\label{sec:level1}Sample Preparation and Electrical Characterization}

Exchange biased spin-valve films consisting of Ta(3)/ NiFe(1.5)/
IrMn(5)/ CoFe(2.5)/ Cu(3)/ CoFe(4)/ NiFe(15)/ Ta(2.5) (thicknesses
in nm) were sputtered onto a Si substrate. To fabricate the
devices, we used a combination of e-beam lithography and reactive
ion etching to produce a thin carbon hardmask. This mask protected
the devices during the Ar ion milling used to etch the spin valve
film. A second e-beam step was used to pattern thick Ta/Au leads
allowing 4-probe device resistance measurements to be made. The
left side of Fig. \ref{Fig1}(a) is a schematic plane view of the
device. There are two important features: a 50 $\mu$m square
nucleation pad and a long (100 $\mu$m), narrow (0.5 $\mu$m)
propagation line. The four gold leads (I+, V+, I- and V-) for
measuring the propagation line resistance are also shown. A
rectangular notch is positioned one-third of the way along the
propagation line (see Fig.\ref{Fig1}(a) right). Different notch
geometries have been fabricated, varying the notch width (W) from
500 nm to 50 nm.

Typical curves of the propagation line resistance as a function of
the applied field (H$_{app}$) are shown in Fig.\ref{Fig1}(b). The
resistance measurements were performed at room temperature using a
standard 4-probe technique. The GMR signal is recorded during two
different field loops. The first, corresponding to the
($-\blacktriangle-$) curve, is a minor loop during which the
magnetization of the CoFe/NiFe bilayer (called free layer in the
following) is completely reversed. The minimum and maximum states
of resistance correspond respectively to the parallel and
antiparallel alignment of the spin-valve magnetizations.
Intermediate plateaus at values corresponding to either 1/3 or 2/3
of the total resistance variation are present during the switching
of the free layer magnetization. These plateaus indicate the
existence of a magnetic configuration in which a DW present in the
free layer is trapped at the notch\cite{Ono98}. In order to check
 the stability of this configuration at remanence, the propagation line
resistance has been recorded during a second field loop: the
($-$o$-$) curve in Fig. \ref{Fig1}(b). After preparing the sample
in a magnetic state where a DW is trapped in the notch, H$_{app}$
is reversed. Note that the resistance remains at an intermediate
level for -20 Oe$<$ H$_{app}<$+ 3 Oe. So in this field range the
DW stays trapped by the notch. To further investigate this
specific magnetic configuration, MFM measurements have been
performed.

\section{\label{sec:level2}Results and Discussion}

Using a commercial Digital Instruments scanning probe microscope
equipped with a standard MESP tip, both topographic and MFM
measurements were performed at room temperature. No external
magnetic field was applied during the MFM scans, and the
tip-sample separation was 18 nm. The topographic images for four
different notch shapes (W= 500, 200, 100 and 50 nm) are denoted
``Topo." in Fig. \ref{Fig2}. The depth of the notch is 150 nm for
all widths except for samples with W=50 nm where proximity effects
in the negative e-beam resist made the notch depth 110 nm. With
the electrical characterization technique previously described, we
have been able to prepare the free layer of our samples in two
different magnetic states at remanence named A and B in the
following (see the resistance levels denoted A and B in Fig.
\ref{Fig1}(b)). The state A is characterized by the presence of a
DW trapped by the notch. Conversely, no DW is present in the state
B. The MFM scans obtained in the states A and B, for samples with
W=500, 200, 100 and 50 nm, correspond respectively to the pictures
labelled MFMA and MFMB in Fig. \ref{Fig2}. Independent of the
notch geometry, dark shapes (resembling a bat) centered on the
notch are present on the MFMA pictures, while two black and white
areas surrounding the notch are observable in the MFMB pictures.

The contrast of MFMB images have been reproduced well using the
micromagnetic model OOMMF\cite{OOMMF} (not shown here). Using the
same model, the micromagnetic configuration and resulting MFM
image at remanence has also been computed in the case of a trapped
DW (state B) for different reverse fields values. Figure
\ref{Fig3} shows the results of these simulations. The absolute
value of the reverse field is increased in 20 Oe increments,
starting from 60 Oe for simulation (i) and ending at 120 Oe for
simulation (iv). Note that the effect of temperature is not taken
into account in these simulations, leading to an overestimation of
the propagation field values. Depending on the reverse field
value, different wall shapes and wall positions can be attained at
remanence. So the simulations predict the existence of multiple
possible trapped DW positions at remanence. But none of these
simulations produces MFM images similar to the observed ``bat"
shape which our magnetoresistance measurements unambiguously
associate with the presence of a trapped DW. Note, however, that
the observed ``bat" shape may be obtained via a superposition of
the different simulated states and of their mirrors. This strongly
suggests that the DW is imaged in different positions during the
same MFM scan. Hence, the observed ``bat" shape likely originates
from DW motion induced by the tip. Such motion is not surprising.
The effect of MFM tips on micromagnetic configuration was widely
studied when the technique was introduced; specifically, DW motion
effects induced by MFM tips were observed by Mamin, et
al.\cite{Mamin89}. In order to confirm this hypothesis we have
performed MFM scans in which we vary the scanning history.

Figure \ref{Fig4} shows the results obtained when a DW is present
in the free layer for devices with W= 500, 200, 100 nm. The
smaller notch case (W= 50 nm) will be discussed later. Three
different scanning procedures have been used; they are named (a),
(b) and (c) in the following. Each of them includes two successive
scans and the pictures are always recorded during the second scan.
The procedure (a) consists in a scan starting from the ``bottom"
of the device and ending at the ``top" of the element immediately
followed by a scan performed in the opposite direction (starting
from the ``top"). During procedure (b) the two successive scans
are performed in the same direction from the ``bottom" to the
``top" part of the device. For procedure (c) the two successive
scans are performed from the ``top" to the ``bottom". The global
trajectory followed by the tip during each procedure is noted by
arrows in Fig. \ref{Fig4}. The MFM pictures recorded using
procedures (a), (b) and (c) are labeled respectively (a), (b) and
(c) in Fig. \ref{Fig4}. If procedure (a) leads to the observation
of the full ``bat" shape as expected\footnote{Indeed, we have used
procedure (a) for the acquisition of pictures shown in Fig.
\ref{Fig2}. Note that (a) is the default procedure currently used
in scanned probe microscopy.}, procedures (b) and (c) generate a
discontinuity during the magnetic signal recording. Only the top
(bottom) wing of the ``bat" is present in images when procedure
(b) (procedure (c)) is performed. These three behaviors can be
easily understood by taking the DW motion induced by the MFM tip
into account. During the first scan of procedure (a), the tip
leaves the DW in a position bordering the top of the notch, and
then during the second scan, the tip, moving from top-to-bottom,
displaces the DW through the notch. So the DW is imaged in all of
the trapping positions induced by the notch, and this leads to the
complete ``bat" observation. At the end of the first scan of
procedure (b), the tip leaves the DW in a position bordering the
top of the notch. During the second scan, the MFM tip is unable to
image the DW in the bottom position, simply because the domain
wall is situated in the top position. So the bottom wing of the
``bat" is missing. When the tip approaches close enough to the
domain wall, the wall is attracted in an irreversible way, and
this leads to the discontinuity in the magnetic signal observed in
(b). Then the DW experiences the trapping positions situated on
the tip trajectory. This leads to the imaging of the top part of
the ``bat". If the two scanning directions of procedure (b) are
reversed, we expect that the top wing of the ``bat" will be
missing, as observed in the images recorded using procedure (c).
The same experiments were performed on the smallest notch devices
(W= 50 nm), but the results obtained were not reproducible from
one device to another. We believe that this is due to weaker
pinning in the shallower notch depth realized at this dimension.


In conclusion, stripe shaped spin-valves including a sub-micron
notch have been fabricated in order to investigate trapped DWs.
Combining micromagnetic simulations, transport, and MFM
measurements, we have demonstrated that multiple stable states can
be created by the notch. Furthermore, the shape and the stability
of the wall appears to be dependent on the trapping position.
These results should be taken into account when dealing with
spin-dependent electronic devices involving trapped DWs because
the magneto-electronic properties can change drastically from one
stable position to another.

\begin{acknowledgments}
We wish to acknowledge M. Hehn and Z. Z. Bandic for helpful
discussions.
\end{acknowledgments}

\pagebreak

REFERENCES


\pagebreak

FIGURE CAPTIONS

\begin{figure}[hbt]
\caption{\label{Fig1} (a) Schematic representation of the device.
(b)  Resistance as a function of the applied magnetic field
recorded at room temperature for two different field loops
(W=0.5$\mu m$). ($-\blacktriangle-$) corresponds to a minor loop
in which the magnetization of the free layer is completely
switched. During the ($-$o$-$) loop, a DW is trapped in the notch
and then stabilized in zero field.}

\caption{\label{Fig2}Topographic (Topo.) and MFM (MFMA, MFMB)
pictures of notches for different geometries. The MFMA pictures
correspond to the trapped DW state, while the MFMB images have no
DW present.}

\caption{\label{Fig3} Micromagnetic simulations of a trapped DW at
remanence for different reverse field values. The absolute values
of the reverse fields are respectively equal to 60, 80, 100, 120
Oe for the simulations (i), (ii), (iii) and (iv).}

\caption{\label{Fig4}Schematic representation of the three
scanning procedures and MFM pictures of the trapped DW
configurations recorded in using these procedures.}

\end{figure}
\pagebreak

\begin{center}
\includegraphics{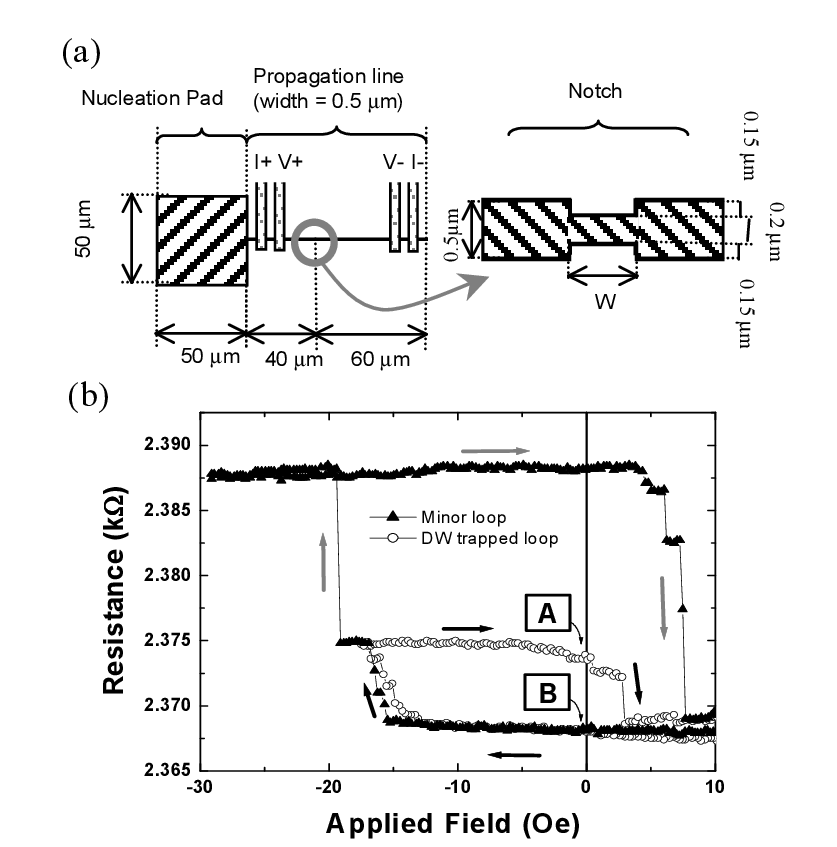}

Fig. 1 Lacour et al.
\end{center}
\pagebreak

\begin{center}
\includegraphics{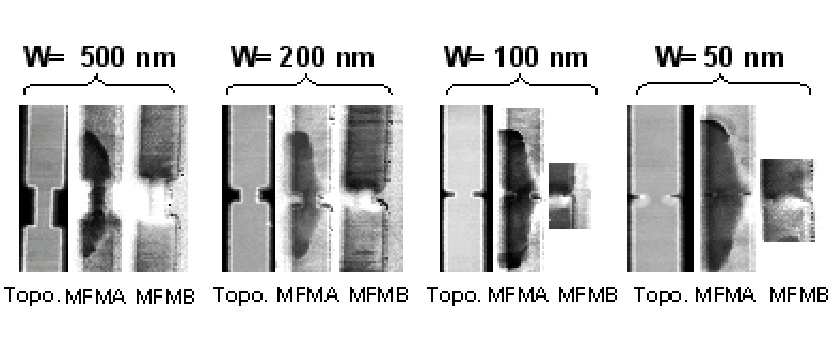}

Fig. 2 Lacour et al.
\end{center}
\pagebreak

\begin{center}
\includegraphics{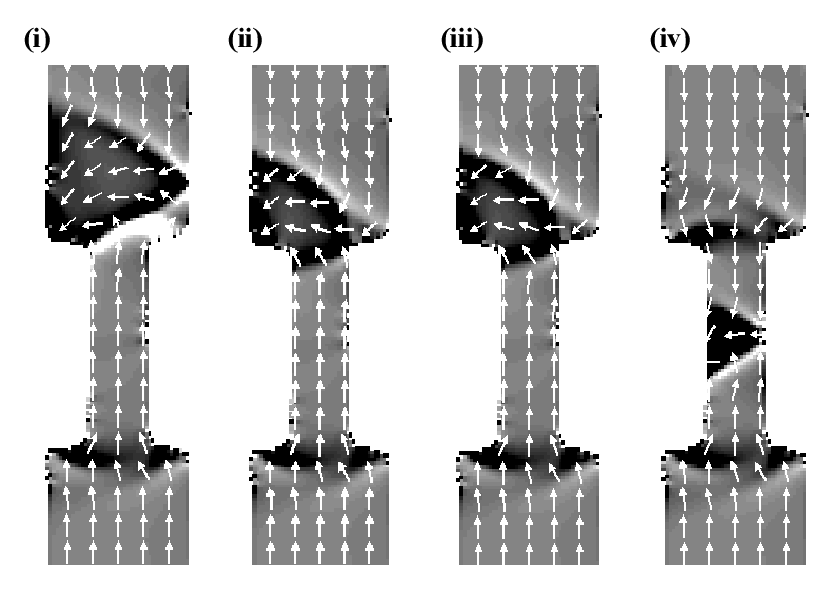}

Fig. 3 Lacour et al.
\end{center}
\pagebreak

\begin{center}
\includegraphics{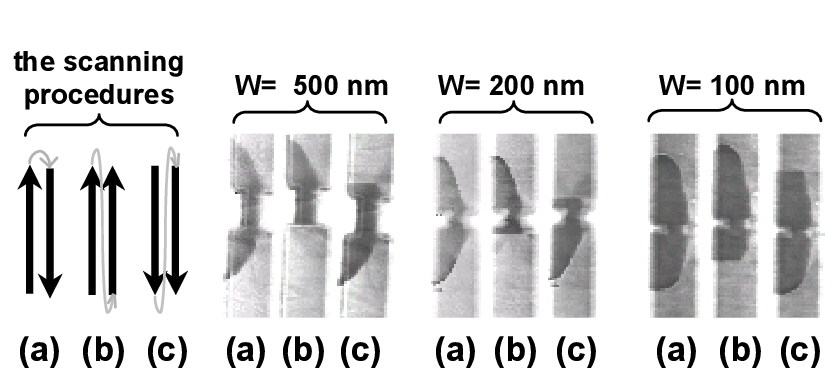}

Fig. 4 Lacour et al.
\end{center}
\pagebreak

\end{document}